\documentclass[secnumarabic,superscriptaddress,amssymb, 10pt, nobibnotes,aps, prl]{revtex4}

\usepackage{graphicx, units, color}
\usepackage{epsfig}

\begin{document}

\title{Deterministic coupling of a single silicon-vacancy color center to a photonic crystal cavity in diamond}%

\author{Janine Riedrich-M\"oller}
\affiliation{Universit\"at des Saarlandes, Fachrichtung 7.2 (Experimentalphysik), Campus E 2.6, 66123 Saarbr\"ucken, Germany}

\author{Carsten Arend}
\affiliation{Universit\"at des Saarlandes, Fachrichtung 7.2 (Experimentalphysik), Campus E 2.6, 66123 Saarbr\"ucken, Germany}

\author{Christoph Pauly}
\affiliation{Universit\"at des Saarlandes, Fachrichtung 8.4 (Materialwissenschaft und Werkstofftechnik), Campus D 3.3, 66123 Saarbr\"ucken, Germany}

\author{Frank M\"ucklich}
\affiliation{Universit\"at des Saarlandes, Fachrichtung 8.4 (Materialwissenschaft und Werkstofftechnik), Campus D 3.3, 66123 Saarbr\"ucken, Germany}

\author{Martin Fischer}
\affiliation{Universit\"{a}t Augsburg, Lehrstuhl f\"ur Experimentalphysik IV, 86159 Augsburg, Germany}

\author{Stefan Gsell}
\affiliation{Universit\"{a}t Augsburg, Lehrstuhl f\"ur Experimentalphysik IV, 86159 Augsburg, Germany}

\author{Matthias Schreck}
\affiliation{Universit\"{a}t Augsburg, Lehrstuhl f\"ur Experimentalphysik IV, 86159 Augsburg, Germany}

\author{Christoph Becher}
\email[]{christoph.becher@physik.uni-saarland.de}
\affiliation{Universit\"at des Saarlandes, Fachrichtung 7.2 (Experimentalphysik), Campus E 2.6, 66123 Saarbr\"ucken, Germany}

\title{Deterministic coupling of a single silicon-vacancy color center to a photonic crystal cavity in diamond}%


\begin{abstract}
Deterministic coupling of single solid-state emitters to nanocavities is the key for integrated
quantum information devices. We here fabricate a photonic crystal cavity around a preselected
single silicon-vacancy color center in diamond and demonstrate modification of the emitters internal population dynamics and radiative quantum efficiency. The controlled, room-temperature cavity coupling gives rise to a resonant Purcell enhancement of the zero-phonon transition by a factor of 19, coming along with a 2.5-fold reduction of the emitter's lifetime. 
\end{abstract}

\maketitle

\section{Introduction}

The application of color centers in diamond for quantum information processing and quantum communication enormously profits from coupling to optical cavities \cite{Aharonovich2011,Loncar2013}: the effects range from simple enhancement of the collection efficiency, which would be beneficial e.g. for remote entanglement schemes \cite{Sipahigil2012,Bernien2013} relying on efficient interference of photons from independent emitters, to cavity-enabled schemes like efficient readout of the spin-state \cite{Young2009} or architectures for scalable quantum information processing \cite{Nemoto2013, Nickerson2013}. The most prominent color centers in diamond are the nitrogen-vacancy (NV) and silicon-vacancy (SiV) centers: NV centers \cite{Doherty2013} have been proven to provide all essential requirements for quantum information  processing \cite{Ladd2010} such as storing, processing and transmitting quantum information.  Their success is largely due to the long spin coherence times \cite{Bar-Gill2013} and strongly selective spin-dependent fluorescence. On the other hand,  SiV centers feature very narrow emission lines even at room temperature \cite{Neu2011} which clearly distinguishes them from the broad-band NV center emission. In addition, very recently, evidence for an optically accessible electron spin of the SiV center has been reported \cite{Muller2014} possibly enabling spin-photon quantum interfaces. The coupling of single color centers to optical cavities with very small modal volume, i.e. photonic crystal (PhC) cavities, has been demonstrated for single NV centers \cite{Faraon2012b, Hausmann2013} at cryogenic temperatures necessary to reduce the zero-phonon line (ZPL) broadening. In these experiments the color centers' position within the cavities was completely random and cavity-emitter systems were selected after fabrication for optimum coupling. Here, on the contrary, we investigate the deterministic coupling of single SiV centers to PhC cavities by a tailored fabrication process:  we fabricate monolithic PhC cavities in single crystal diamond around pre-characterized SiV centers, enabling deterministic emitter positioning and controlled alignment of its dipole axis with the cavity electric field. Furthermore, we profit from the unique spectral properties of the SiV center allowing for a demonstration of cavity quantum electrodynamic  effects even at room temperature.

\section{Fabrication and characterization of cavities}
\begin{figure}[t]
	\centering
		\includegraphics[width=4in]{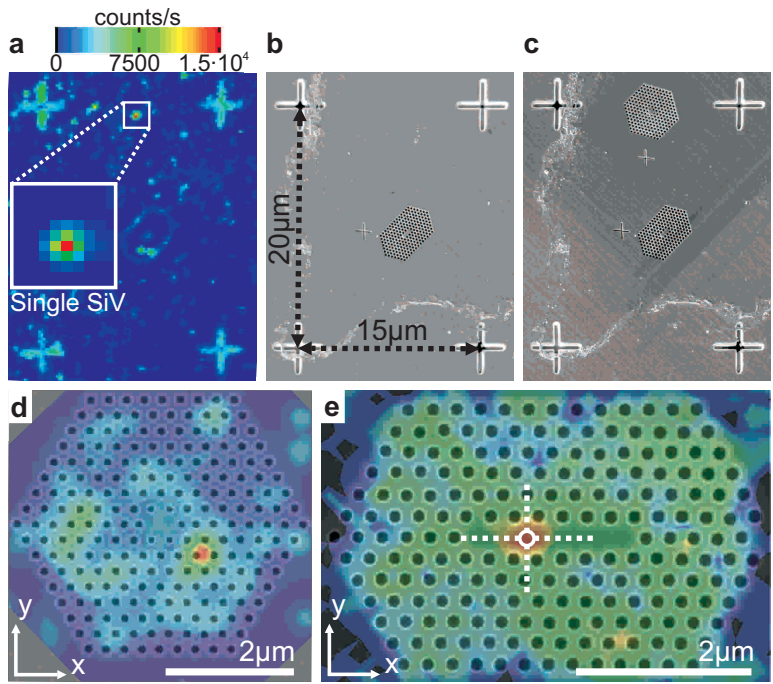}
	\caption{(Color) Deterministic positioning of photonic crystal cavities: (a) Fluorescence scan of the diamond membrane with single SiV centres and positioning markers. (b,c) Scanning electron microscope images (b) before and (c) after patterning a M1-cavity at the SiV center position (same sample region as in (a)). (d,e) Scanning electron microscope images of fabricated (d) M1- and (e) M7-cavity overlapped with  fluorescence scans confirming the presence of single emitters within the cavities.}
	\label{fig:position}
\end{figure}

The PhC consists of a  triangular lattice of air holes with a pitch of $a \approx $ \unit[283]{nm} and a hole radius of $r \approx $ \unit[85]{nm}, that is milled in a free-standing single crystal diamond membrane with a thickness of  $h \approx $ \unit[400]{nm}. A nanocavity is formed by a linear one- or seven-hole defect in  the photonic lattice, labeled M1- or M7-cavity, respectively. 
The single crystal diamond slab (orientation of the top face is (001)) is grown by chemical vapor deposition  on a silicon substrate with iridium/yttria-stabilized zirconia buffer layers \cite{Gsell2004}. The diamond growth parameters are adjusted such that a small amount of silicon is incorporated in the diamond yielding on average 2 single SiVs/\unit[$20\times20$]{$\mu$m$^2$} in the final membrane. To obtain a free-standing membrane, we remove the sacrificial substrate and thin out the diamond to a desired thickness of approximately \unit[400]{nm} by reactive ion etching in an oxygen plasma \cite{Riedrich-Moller2012}. To locate single emitters, markers are etched into the membrane. The single SiV centers and positioning markers are observed in the photoluminescence (PL) scan (Fig. \ref{fig:position}(a)). For obtaining PL we use a confocal microscope setup (objective: 100$\times$, numerical aperture 0.8) with a continuous wave excitation laser (\unit[660]{nm}) and a spectrometer with liquid-nitrogen cooled CCD camera. A Hanbury Brown and Twiss interferometer enables measurements of the photon intensity correlation function and verification of single emitters. All experiments are performed
at room temperature.
 The  alignment markers are subsequently used to pattern a photonic crystal cavity around a single emitter using focused ion beam (FIB) milling (\unit[30]{keV}-Ga$^{+}$ ions) \cite{Riedrich-Moller2012}.  Figures \ref{fig:position}(b,c) show scanning electron microscope  images before and after the cavity structuring. After annealing at \unit[1,000]{$^{\circ}$C} and cleaning the sample in a 1:1:1 boiling mixture of nitric, sulfuric, perchloric acid, PL scans reveal the presence of single emitters within the cavity.  Overlapping the PL scans with the scanning electron microscope images of the device (Fig. \ref{fig:position}(d,e)), we can estimate the  positioning accuracy  to be as small as one lattice constant. 
Whereas in figure \ref{fig:position}(d) the emitter is displaced from the cavity center but still residing in the photonic lattice, the emitter of figure \ref{fig:position}(e) is perfectly located in the central cavity region. 
The positioning accuracy is mainly limited by the resolution (\unit[$\approx500$]{nm}) of the fluorescence scans.

\begin{figure}[t]
	\centering
		\includegraphics[width=4.5in]{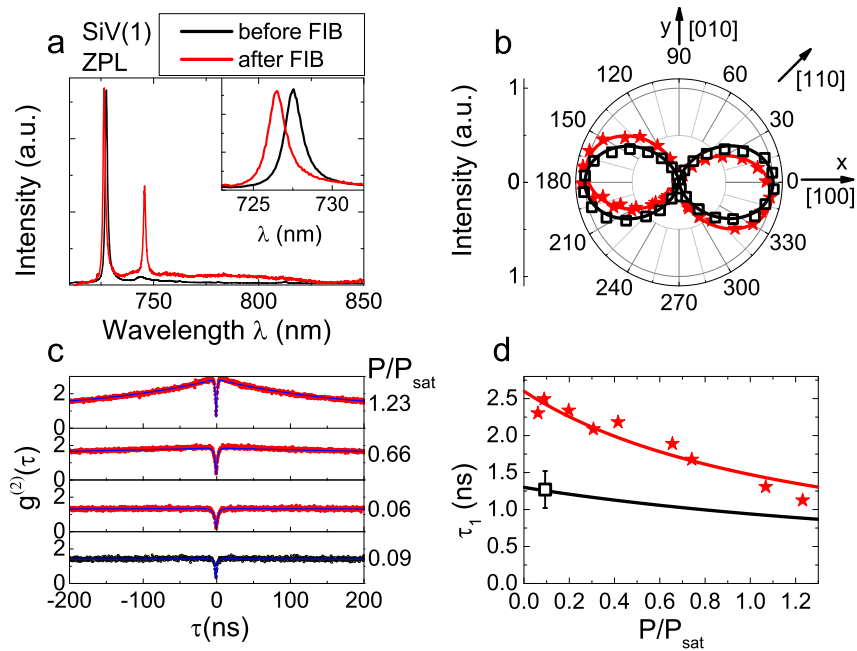}
	\caption{(Color) Emission properties upon structuring: (black: before FIB, red: after FIB) (a) PL spectra of a single SiV(1) center and (b) polarization measurements taken before  and after  the FIB milling of the M1-cavity (Fig. \ref{fig:position}(d)). The inset in (a) shows a blue-shift by \unit[1]{nm} of the SiV(1) ZPL at \unit[727.5]{nm} upon structuring. (c)  Intensity correlation measurements $g^{(2)}$ of the SiV(1) ZPL for selected excitation powers $P$ normalized to the saturation power $P_{\mathrm{sat}} =$ \unit[1.0]{mW} prior to FIB and $P_{\mathrm{sat}} =$ \unit[0.76]{mW} after FIB. Solid lines: Fit curves of equation (\ref{eq:g2}).  (d) Parameter $\tau_1$ extracted from the $g^{(2)}(\tau)$ fits for various excitation powers. }
	\label{fig:M1}
\end{figure}

For application of the cavity alignment technique with respect to a pre-characterized emitter, it is crucial, that its spectral position, linewidth and dipole orientation are persistent upon fabrication. 
Figures \ref{fig:M1}(a) and \ref{fig:tuning}(a) show the PL spectra taken before and after the FIB milling of the M1- and M7-cavity around  single emitters named SiV(1) and SiV(2), respectively. 
In both cases, the ZPL  at $\lambda_{\mathrm{SiV(1)}}=$ \unit[727.5]{nm} and $\lambda_{\mathrm{SiV(2)}}=$ \unit[726.0]{nm} blue or red shifts by several nanometers upon structuring 
(insets in Figs. \ref{fig:M1}(a), \ref{fig:tuning}(a)). The deviation of the ZPL wavelengths from the standard value of $\lambda_{\mathrm{ZPL}}  =  $ \unit[738]{nm} is due to strain in the sample blue-shifting the ZPL \cite{Neu2011a}. The ZPL wavelength shift upon fabrication can be attributed to a local relaxation of the crystal strain due to the air hole patterning.
In contrast, the change of the  ZPL linewidths is $<10\%$, indicating that almost no additional broadening mechanisms are induced by the structuring process. 
Besides the original emission lines new ZPLs at \unit[745.7]{nm} and  \unit[739.9]{nm} appear in the M1- and M7-cavity spectra after  FIB milling, respectively. Etching holes in the diamond introduces  additional vacancies that form new SiV centers after the subsequent annealing step.  The creation of new SiV centers is observed in almost every fabricated PhC.

Recent experiments \cite{Hepp2014, Rogers2014} show strong evidence of a $\langle 111 \rangle$ orientation of the SiV center and its dominant dipole moment in the diamond lattice. Observing emission from the top of the sample, a $\langle 111 \rangle$ oriented dipole appears in a $\langle 110 \rangle $ direction due to projection onto the $(001)$ sample plane.
Placing a polarization analyzer in the detection path and detecting the PL signal by a spectrometer as a function of the azimuthal angle $\Phi$ we determine the orientation of the emission dipole within the (001) diamond plane. The crystallographic axes of the diamond sample are specified via electron backscatter diffraction. 
Our measurements (Figs. \ref{fig:M1}(b), \ref{fig:tuning}(c), \ref{fig:tuning1015}(a)) show an effective linear polarization of the SiV center along $\langle 100 \rangle$ and $\langle 110 \rangle$ directions within the $(001)$-plane, which is conform with previous measurements of single SiVs in strongly strained nanocrystals \cite{Neu2011a}. Theoretical simulations of the SiV center electronic transitions \cite{Hepp2014} reveal that under the influence of strain the polarization orientation rotates away from $\langle 110 \rangle$ towards $\langle 100 \rangle$, confirming the experimental observations. We choose a coordinate system  such that an azimuthal angle $\Phi = 0^\circ$ (90$^\circ$) corresponds to a linear polarization along the  $x$- ($y$-) axis, respectively. The SiV(1) emission is linearly polarized along the $x$-axis (Fig. \ref{fig:M1}(b)), which is persistent upon FIB milling within the detection precision of $10^\circ$, confirming the preservation of dipole orientation upon PhC fabrication.

\section{Cavity coupling effects}

In the following we demonstrate different effects of emitter-cavity coupling such as tuning cavity modes across the entire emission spectrum of SiV centers as well as inhibition and enhancement of spontaneous emission (SE). In a homogeneous medium the SE rate $\gamma_0$ of a color center is given as the sum of radiative decay rates into ZPL, $\gamma_{\mathrm{ZPL}}$, and phonon side bands,  $\gamma_{\mathrm{PSB}}$, and the non-radiative decay rate  $\gamma_{\mathrm{nr}}$:  $\gamma_0 = \gamma_{\mathrm{ZPL}} + \gamma_{\mathrm{PSB}}+ \gamma_{\mathrm{nr}}$. Coupling a PhC microcavity mode to a particular transition $i$, the decay rate $\gamma_i$ is enhanced by the Purcell factor $F_{\mathrm{cav}}$ whereas other radiative transitions $\gamma_{j\neq i} $ are inhibited by a factor $F_{\mathrm{PhC}}<1$ due to the reduced local density of states in the photonic band gap \cite{Fujita2005}: $\gamma_\mathrm{cav} = F_{\mathrm{cav}}  \gamma_i + F_{\mathrm{PhC}}\sum_{j \neq i} \gamma_j + \gamma_\mathrm{nr}$. $F_{\mathrm{cav}}$ is given by:
\begin{eqnarray}
F_{\mathrm{cav}} &=& F_P \; \frac{1}{1 + 4 Q^2(\frac{\lambda_i}{\lambda_c}-1)^2} \left\langle{\vec{\epsilon}(\vec{r}_i) \cdot \vec{\mu}_i} \right\rangle ^2{\left|\vec{\epsilon}(\vec{r}_i) \right|^2}  \\
&=& F_P \, R_{\lambda} \, R_{\mu}\, R_{r}
\label{eq:Purcell_korr}
\end{eqnarray}
Here $\vec{\epsilon}$ is the cavity electric field $\vec{E}$ normalized to the field maximum $\vec{\epsilon}(\vec{r}_i) = \vec{E}(\vec{r}_i)/\mathrm{max} (\vec{E}(\vec{r}))$ and $\vec{\mu}_i$ is the normalized dipole moment of transition $i$. The factors $R_r={\left|\vec{\epsilon}(\vec{r}_i) \right|^2}$ and $R_{\lambda}={(1 + 4 Q^2(\frac{\lambda_i}{\lambda_c}-1)^2)^{-1}}$  account for the spatial misalignment of the emitters' position and the spectral detuning of the emitter's wavelength $\lambda_i$ from the cavity mode $\lambda_c$ with a quality factor $Q$, respectively. The term  $R_{\mu}=\left\langle{\vec{\epsilon}(\vec{r}_i) \cdot \vec{\mu}_i} \right\rangle ^2 $  accounts for the alignment of the dipole emission with the cavity electric field. For a dipole resonant with the cavity, perfectly positioned and oriented with the cavity electric field, the ideal Purcell factor is $F_P = \frac{3}{4 \pi^2}\frac{Q}{V}\left(\frac{\lambda_c}{n}\right)^3$ \cite{Purcell1946}, where $V$ is the mode volume and $n = 2.4$ the refractive index of diamond. This model assumes that the cavity linewidth $\Delta \lambda_c = \lambda_c/Q$ is larger than the linewidth of the coupled transition $\Delta \lambda_i$ which is a valid assumption for the narrow-band SiV centers and our low-Q factor cavities. In the case of a broad-band emitter, such as the NV center, one would have to include a reduced spectral overlap and cavity feeding effects for non-resonant transitions \cite{Albrecht2013}. Cavity coupling effects as discussed here are detected in our experiments by investigating modifications of the emitters' internal population dynamics.

\begin{figure}[t]
	\centering
				\includegraphics[width=5in]{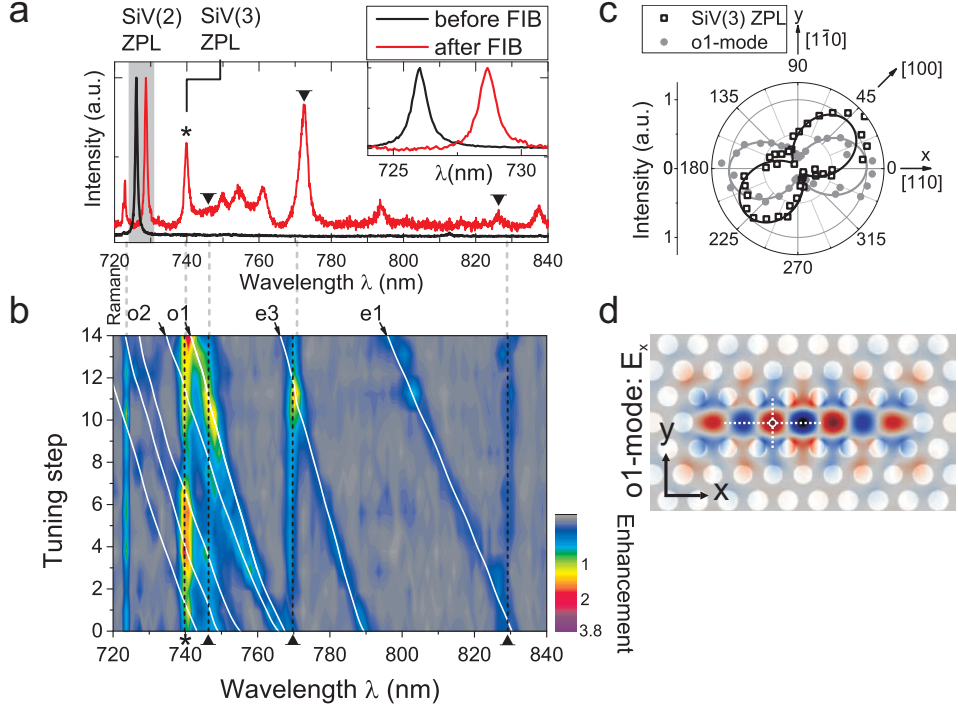}
	\caption{(Color) Spectral tuning of photonic crystal cavity modes: (a) PL spectra taken before (black) and after (red) the fabrication of the M7-cavity (Fig. \ref{fig:position}(e)) around a single SiV(2) center. The inset shows a close up of the SiV(2) ZPL at \unit[726.0]{nm}, that red shifts by \unit[2.6]{nm} 
	upon FIB etching. By the milling process, a new SiV(3) center (ZPL at \unit[739.9]{nm} marked by $\star$, PSB marked by $\blacktriangledown$) is created. At wavelengths \unit[750-830]{nm} M7-cavity modes are visible. (b) 
Cavity tuning of modes $e1$, $e3$, $o1$ and $o2$ into resonance with the SiV(3) ZPL ($\star$) and PSB ($\blacktriangle$). On resonance with the $o1$-mode, the SiV(3) ZPL intensity is enhanced  by a factor of 3.8.  (c) Polarization measurement of the $o1$-mode (\textcolor[rgb]{0.5,0.5,0.5}{$\bullet$}) and SiV(3) ZPL (\textbf{\tiny{$\square$}}\normalsize). (d) Simulated $E_x$-field of $o1$-mode. The emitter's position is marked by a white circle. }
	\label{fig:tuning}
\end{figure}

To verify the presence of a single emitter and to deduce its population dynamics, we measure the intensity correlation function $g^{(2)}$ (e.g. Fig. \ref{fig:M1}(c)) for various excitation powers using a Hanbury Brown and Twiss setup. 
 Modeling the SiV center as a three-level system including a shelving state, the data is fitted by \cite{Neu2011}:
\begin{equation}
g^{(2)} (\tau) = 1-(1+a) \: e^{-\mid \tau \mid /\tau_1} + a \: e ^{-\mid \tau \mid /\tau_2}
\label{eq:g2}
\end{equation}
Comparing the fit parameters at different excitation powers with a three-level rate equation model yields level populations and transition rates. Together with an additional measurement of the saturation count rate this allows for an estimation of the emitter's radiative quantum efficiency \cite{Neu2012a}: $\eta_{\mathrm{qe}} = \gamma_{\mathrm{rad}} /(\gamma_{\mathrm{rad}} + \gamma_{\mathrm{nr}})$ where $\gamma_{\mathrm{rad}}$ is the sum of all radiative rates. 

We first demonstrate inhibition of SE for the SiV(1) center placed in a photonic bandgap structure but spatially and spectrally decoupled from the cavity mode (Fig. \ref{fig:position}(d)). The fit parameters for $\tau_1$ are shown in figure \ref{fig:M1}(d) ($\tau_2$-, $a$-fits see Supplemental Material).  By extrapolating $\tau_1$ for vanishing pump powers, we infer a lifetime of the excited state of $\tau_{1,\mathrm{PhC}}^0 = $ \unit[2.6]{ns}, when the SiV(1) center is placed in the photonic lattice compared to $\tau_{1,\mathrm{bulk}}^0 = $ \unit[1.3]{ns} for the unstructured case. From the rate equation model we also extract the rate $k_{21}$ for the transition from the excited state $|2\rangle$ to the ground state $|1\rangle$ which is subject to SE modifications. Here, $k_{21}^{\mathrm{PhC}} = (\tau_{1,\mathrm{PhC}}^0)^{-1}$ and $k_{21}^{\mathrm{bulk}} = (\tau_{1,\mathrm{bulk}}^0)^{-1}$ resulting in an inhibition factor of 2. It is important to note that the placement of an emitter into a thin dielectric slab, i.e. the unpatterned diamond membrane, already could give rise to SE modification. For our parameters, $h/\lambda \approx 0.54$, however, one does only expect very small effects \cite{Khosravi1992, Brueck2000} such that we can treat the unpatterned membrane as bulk material.
 Finite-difference time-domain simulations of single dipoles at different positions within the unit cell of a triangular PhC lattice predict an inhibition factor $F_\mathrm{PhC} = 0.25$ corresponding to an average lifetime increase by a factor of 4 (for simulations details see Supplemental Material). 
The difference to the observed SE inhibition factor is due to the finite radiative quantum efficiency of SiV centers: from the observed SE rate of SiV(1) in bulk and in the PhC bandgap we can infer theoretical quantum efficiencies $\eta_{\mathrm{qe, bulk}}^{\mathrm{th}} = 0.66$ and $\eta_{\mathrm{qe, PhC}}^{\mathrm{th}} = 0.33$, respectively (see Supplemental Material). An independent measurement of $\eta_{\mathrm{qe}}$ from saturation and population dynamics yields $\eta_{\mathrm{qe, bulk}}^{\mathrm{exp}} = 0.63\pm 0.13$ and $\eta_{\mathrm{qe, PhC}}^{\mathrm{exp}} = 0.18 \pm 0.04$ (see Supplemental Material) which is in reasonable agreement.

As a prerequisite for the demonstration of SE enhancement, we show next that the cavity modes can be tuned across the entire spectrum of the SiV(3) center (Fig. \ref{fig:tuning}(a,b)): 
At wavelengths \unit[750-830]{nm}, modes of the M7-cavity are visible in the  spectrum (Fig. \ref{fig:tuning}(a)), red-detuned with respect to the ZPL at \unit[739.9]{nm} of the newly created SiV(3) center.
To blue tune the cavity modes, we use a digital etching technique by heating the sample in an oxygen atmosphere \cite{Riedrich-Moller2012}. In total, the cavity modes are blue shifted up to \unit[45.6]{nm} with a mean tuning rate of \unit[1.6]{nm} per oxidation step. 
As expected we observe a SE enhancement when different modes are tuned into resonance with the ZPL at \unit[739.9]{nm}. We detect a maximum intensity enhancement by a factor of 3.8, when shifting the $o1$-mode with a quality factor of $320\pm75$ and a mode volume of $1.3\, (\lambda/n)^3$ in resonance with the SiV(3) ZPL (Fig. \ref{fig:tuning}(c)). Interestingly, there are a number of additional resonances appearing in the SiV(3) side band at  \unit[828.6]{nm}, \unit[769.0]{nm} and \unit[746.3]{nm}. As cavity enhancement is only effective for narrow-band transitions comparable to the cavity bandwidth (\unit[2.3]{nm}) we tentatively assume that the coupled transitions are electronic transitions, in addition to the ZPL. This assumption is supported by recent experimental observations \cite{Neu2012} and theoretical models \cite{Gali2013} of the SiV center.
The ZPL intensity enhancement of 3.8 is comparable to previous ensemble measurements of SiV centers \cite{Riedrich-Moller2012}. 
The ideal Purcell enhancement that can be observed from the PL spectrum is given as the ratio of radiative rates on (off) resonance with the cavity mode, which are enhanced (inhibited) by Purcell coupling (photonic bandgap effect): $I_{\mathrm{PL}} = F_{\mathrm{cav}}\gamma_{\mathrm{ZPL}}/F_{\mathrm{PhC}}\gamma_{\mathrm{ZPL}}$. The ideal value $I_{\mathrm{PL}} = 75$ is reduced by non-optimal dipole orientation $R_{\mu} \approx 0.25$ (Fig. \ref{fig:tuning}(c)) and non-ideal positioning of the SiV(3) center $R_r \approx 0.25$ (Fig. \ref{fig:tuning}(d)). 
Note that deviations along the $z$-axis have a minor impact: Displacing the emitter by \unit[150]{nm} out of the slab center changes $R_r$ by less than 10\%. With equation (\ref{eq:Purcell_korr}) and $R_{\lambda}\approx 1$ we get $F_{\mathrm{cav}} = 1.17$ and $I_{\mathrm{PL}} = 4.7$, in good agreement with the observed PL enhancement of 3.8.

\begin{figure}[t]
	\centering
		\includegraphics[width=4.7in]{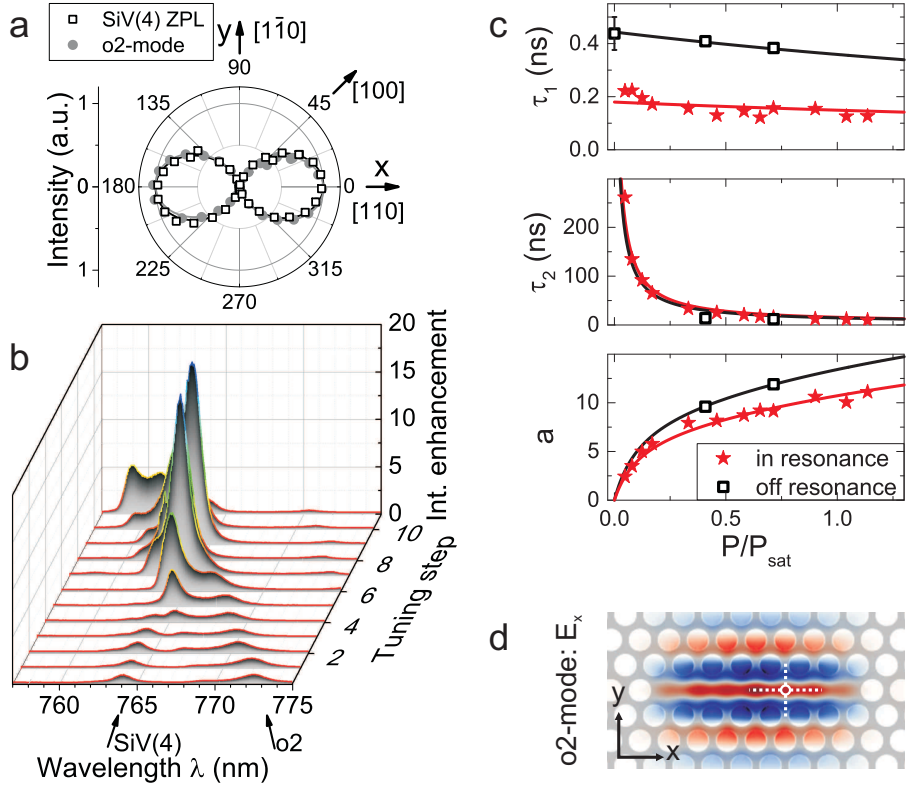}
	\caption{(Color) Purcell enhancement of the SiV(4) ZPL: (a) Polarization polar plot of SiV(4) ZPL (\textbf{\tiny{$\square$}}\normalsize) and the M7-cavity $o2$-mode (\textcolor[rgb]{0.5,0.5,0.5}{$\bullet$}).  (b) Tuning spectra of $o2$-mode: In resonance with the SiV(4) ZPL an intensity enhancement of 19 is observed. (c) On (\textcolor[rgb]{1,0,0}{$\star$}) and off (\textbf{\tiny{$\square$}}\normalsize) resonance with the cavity,  the parameters $\tau_1$, $\tau_2$, $a$ are determined by fitting $g^{(2)}(\tau)$ function measured at various excitation powers $P$ with equation (\ref{eq:g2}). $P$ is normalized to the saturation power  $P_{\mathrm{sat,on}} = $ \unit[0.89]{mW} and $P_{\mathrm{sat,off}} = $ \unit[0.98]{mW} on- and off-resonant with the cavity. Solid lines: theoretical power dependance of $\tau_1$, $\tau_2$, $a$.  (d) $E_x$-field of the $o2$-mode. The emitter's position is marked by a white circle.}
	\label{fig:tuning1015}
\end{figure}

Finally, we demonstrate a large improvement in coupling efficiency by deterministic relative positioning of the cavity and the emitter. To this end, we fabricate a second M7-cavity at the position of a single SiV(4) center with its polarization aligned with the cavity field (Fig. \ref{fig:tuning1015}(a)). Tuning the cavity mode $o2$ with a $Q$-factor of $430\pm150$ and a mode volume of 1.7 ($\lambda/n$)$^3$  in resonance, we observe a PL enhancement by a factor of 19 (Fig. \ref{fig:tuning1015}(b)). 
To gain more insight into the dynamics of the SiV(4) center, we measure the $g^{(2)}$ function for different excitation powers, when the cavity mode is on- and off-resonance with the ZPL. The data is fitted by equation (\ref{eq:g2}). In figure \ref{fig:tuning1015}(c) the parameters $\tau_1$, $\tau_2$, $a$ are shown for various excitation powers $P$. 
For vanishing excitation power, we infer on/off-resonance lifetimes 
$\tau^0_{1, \mathrm{on}} =$ \unit[$180\pm50$]{ps} and $\tau^0_{1, \mathrm{off}} =$ \unit[$445\pm20$]{ps}, respectively. To confirm the $\tau^0_1$ findings, we additionally measure the off-resonance lifetime of the SiV(4) center using a femtosecond Ti:sapphire laser (Spectra Physics Tsunami, \unit[703]{nm}, \unit[80]{MHz} repetition rate) for excitation. The spectrally filtered ZPL signal is detected by an avalanche photodiode (timing jitter \unit[296]{ps}).  We find a lifetime of the excited state of \unit[$440\pm60$]{ps}, which is conform with the $\tau^0_{1, \mathrm{off}}$ time constant deduced by the $g^{(2)}$ measurements. 

The  reduction of the lifetime coming along with the increase in PL are related to the Purcell enhancement of the ZPL transition when coupled to the PhC cavity. The ideal Purcell factor $F_P = 19.2$ determined by the mode volume and quality factor of the $o2$ mode is reduced to $F_{\mathrm{cav}} \approx 5$ when taking into account the spatial  mismatch $R_r = 0.4$ due to non-ideal emitter positioning within the cavity (Fig. \ref{fig:tuning1015}(d)) as well as an inclination angle of 35.3$^\circ$ between the SiV(4) dipole moment along the $\langle 111 \rangle$ crystal axis and the cavity $E_x$-field in the (001) plane, the dominant electric component of the $o2$ mode, yielding $R_{\mu}   = 0.67$. Here, we assume a perfect spectral overlap $R_{\lambda} = 1$ reached  by tuning the  cavity  mode with a linewidth of \unit[1.7]{nm} into resonance with the narrow SiV emission with a linewidth of \unit[1.25]{nm}. The enhancement in PL due to Purcell coupling is then expected to be $I_{\mathrm{PL}} = F_{\mathrm{cav}}\gamma_{\mathrm{ZPL}} / F_{\mathrm{PhC}} \gamma_{\mathrm{ZPL}}\approx 5/0.25 = 20$ which is in very good agreement with the experimentally observed factor of 19. Again, the difference to the SE lifetime modification is due to the non-unity quantum efficiency of the SiV center. By tuning the cavity mode into resonance with the ZPL, cavity coupling only affects the radiative decay rates into the ZPL, $\gamma_{\mathrm{ZPL}}$, whereas the side band emission, $\gamma_{\mathrm{PSB}}$, and the non-radiative decay, $\gamma_{\mathrm{nr}}$, remain unchanged. Thus, the total decay rates on/off resonance are: $\gamma_{\mathrm{cav}}  = F_{\mathrm{cav}} \gamma_{\mathrm{ZPL}}+ F_{\mathrm{PhC}} \gamma_{\mathrm{PSB}} +\gamma_{\mathrm{nr}} $  and $\gamma_{\mathrm{PhC}}  = F_{\mathrm{PhC}} ( \gamma_{\mathrm{ZPL}}+ \gamma_{\mathrm{PSB}}) +\gamma_{\mathrm{nr}} $, respectively. By taking the measured rates $\gamma_{\mathrm{cav}} = k_{21}^{\mathrm{cav}} = $ \unit[5,238]{MHz}  and $\gamma_{\mathrm{PhC}}  = k_{21}^{\mathrm{PhC}} = $ \unit[1,932]{MHz} (see Supplemental Material) and assuming a branching ratio $\gamma_{\mathrm{ZPL}} :\gamma_{\mathrm{PSB}} = 4:1$ \cite{Neu2011} we can infer transition rates $\gamma_{\mathrm{ZPL}} = ($\unit[1.44]{ns}$)^{-1}$, $\gamma_{\mathrm{PSB}} = ($\unit[5.75]{ns}$)^{-1}$ and $\gamma_{\mathrm{nr}} = ($\unit[583]{ps}$)^{-1}$.
This in term allows for the calculation of the underlying quantum efficiencies on resonance $\eta_{\mathrm{qe, cav}}^{\mathrm{th}} = 0.67$, off resonance $\eta_{\mathrm{qe, PhC}}^{\mathrm{th}} = 0.11$ and a hypothetical value for an emitter in unstructured bulk material $\eta_{\mathrm{qe, bulk}}^{\mathrm{th}} = 0.34$.

To confirm these results, we independently determine the quantum efficiencies $\eta_{\mathrm{qe, cav}}^{\mathrm{exp}} = 0.49\pm0.04$ and $\eta_{\mathrm{qe, PhC}}^{\mathrm{exp}} = 0.08\pm0.01$  of the single SiV(4) center from independent saturation measurements and population dynamics (see Supplemental Material).  The experimental findings are in very good agreement with the values calculated from the transition rates. The increase in the quantum efficiency unambiguously proves the impact of the cavity coupling on the emission properties of the single emitter. On resonance, a fraction of 63\% of the total decay rate and 98.8\% of the radiative emission are channeled into the cavity mode.

\begin{figure}[t]
	\centering
		\includegraphics[width=3in]{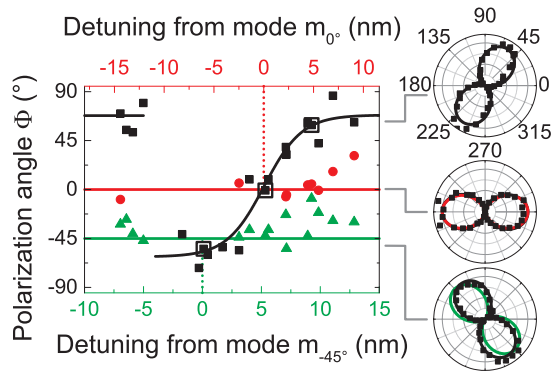}
	\caption{(Color) Continuous polarization control: Polarization emission angle $\Phi$ of the SiV phonon side band (\tiny$\blacksquare$\normalsize) as a function of the detuning from the cavity modes $m_{0^{\circ}}$ and $m_{-45^{\circ}}$ with polarization angles $\Phi = 0^{\circ}$ (\textcolor[rgb]{1,0,0}{$\bullet$}) and $\Phi = -45^{\circ}$ (\textcolor[rgb]{0,0.5,0}{$\blacktriangle$}), respectively. The solid lines are a guide to the eye. On the right: Polar plots of SiV phonon side band (\tiny$\blacksquare$\normalsize) and of cavity modes (red/green solid line) for selected detunings (\footnotesize$\Box$\normalsize).}
	\label{fig:polarization}
\end{figure}

Beyond the change of the SE  the emitter's polarization  can be controlled by cavity-coupling, as  shown previously e.g. for quantum dots weakly coupled to a PhC cavity \cite{Gallardo2010}. Here, we consider a transition in the side band region of the SiV center linearly polarized along the azimuthal angle $\Phi = 60^{\circ}$. By progressively tuning two cavity modes $m_{0^{\circ}}$ and $m_{-45^\circ}$ with polarization angles $\Phi = 0^{\circ}$ and $\Phi = -45^{\circ}$ into resonance with the SiV phonon side band, we continuously change its  polarization emission angle $\Phi$ from $+60^{\circ}$ to $-45^{\circ}$ (Fig. \ref{fig:polarization}). For large detunings the original polarization state is restored. The precise control of the linearly polarized emission paves the way for polarization controlled single photon emitters.

\section{Conclusions}

In summary, we deterministically fabricated PhC cavities around pre-characterized single SiV centers in a single crystal diamond membrane. Upon patterning of the material, the SiV linewidth and dipole orientation are preserved whereas the release of local strain might slightly shift the emission line. The extraordinarily narrow emission lines of SiV centers allow for the demonstration of cavity quantum electrodynamic effects even at room temperature: we observe both inhibition of spontaneous emission by the photonic bandgap effect and enhancement due to cavity coupling. As we can precisely determine all parameters contributing to the cavity coupling, predicted and measured emission rates are in very good agreement. Our analysis also allows to trace the modifications of the radiative quantum efficiency: it is enhanced by a factor $>6$ for Purcell coupling on resonance and reduced by a factor $>2$ in case of SE inhibition. Finally, the polarization angle of SiV emission can be controlled by coupling to the cavity modes paving the way for efficient, polarization-controlled single photon sources. 

The radiative properties of the individual emitters investigated here differ a lot: we find a large variation in ZPL emission wavelengths (\unit[$\approx725-760$]{nm}), SE lifetimes (\unit[$0.35-1.3$]{ns}) and quantum efficiencies ($0.34-0.66$ for emitters in bulk material). It is well known that these properties scatter largely for emitters in strained diamond material.  The wavelengths and lifetimes observed here fall well into the range of previously determined values \cite{Neu2011, Neu2012a}. For comparison of quantum efficiencies, to our knowledge only one study  on single SiV centers exists \cite{Neu2012a}, yielding values in the range of 0.003-0.09 for SiV centers in nanodiamonds. It is well known that for emitters hosted in very small dielectric particles the local density of states can be drastically reduced. For the case of a sub-wavelength spherical particle the emission rate is reduced by $\gamma_{\mathrm{sphere}} = \gamma_{\mathrm{bulk}} \frac{1}{n} \left( \frac{3}{2+n^2}\right)^2$ where $n$ is the refractive index \cite{Chew1988, Inam2014}. For diamond $(n=2.4)$ the reduction factor amounts to 0.06. If emitters SiV(1) and SiV(4) of our study were placed in nanodiamonds their quantum efficiencies would reduce to 0.10 and {0.03}, respectively, coinciding perfectly with the range determined in earlier experiments \cite{Neu2012a}. A large scatter in quantum efficiencies is also well known for NV centers \cite{Mohtashami2013, Inam2014}.

Eventually, we propose two measures to improve the cavity coupling experiments reported here: First, much more predictable and reproducible spectral properties can be expected for SiV centers in low-strain material as demonstrated recently \cite{Rogers2013a}. Second, the overlap of the emitter's dipole orientation and the cavity field might be maximized by using material with preferentially oriented single emitters as available lately \cite{Lesik2014,Michl2014}.

\section*{Acknowledgments}

We cordially thank R. Albrecht and L. Kipfstuhl for helpful discussions on the cavity coupling and C. Hepp for discussion on the SiV dipole orientation. Furthermore, we would like to thank J. Schmauch for scanning electron microscopy and K. Kretsch for assistance with the wet chemical etching. In addition, we  thank S. Wolff and C. Dautermann (Nano Structuring Center, University of Kaiserslautern) for sputtering of metal layers and A. Baur and M. Wandt (IMTEK, University of Freiburg) for deep reactive ion etching. This research has been partially funded by the European Community's Seventh Framework Programme (FP7/2007-2013) under Grant Agreement N$^{o}$ 618078 (WASPS). Furthermore, the EU funding for the project AME-Lab (European Regional Development Fund C/4-EFRE 13/2009/Br) for the FIB/SEM is gratefully acknowledged.

\section*{Authors contributions}
J.R-M.  performed the deterministic coupling experiments, fabricated the photonic crystals and processed the diamond membranes,  analyzed the data  and carried out the numerical modeling of the structures.   C.P., J.R-M. and F.M. performed FIB milling. C.A. and J.R-M. performed the lifetime measurements. M.F., S.G. and M.S. developed the chemical vapor deposition growth process for the diamond films on iridium buffer layers.  J.R-M. and C.B. conceived and designed the experiments. J.R-M. and C.B. wrote the manuscript. All authors discussed the results and commented on the manuscript.

%
%


\end{document}